\documentclass[twocolumn]{aastex702}

\usepackage{amsmath}

\newcommand{\sigr}{\sigma_{\rho/\rho_0}}

\newcommand{\sign}{\sigma_{N/N_0}}

\newcommand{\sigrB}{\sigma_{\rho/\rho_0\mathrm{,Brunt}}}
\newcommand{\sigrBdn}{\sigma_{\rho/\rho_0\mathrm{,Brunt,denoised}}}

\newcommand{\ptwod}{P_{\mathrm{2D}}}
\newcommand{\pthreed}{P_{\mathrm{3D}}}
\newcommand{\rtohalf}{\mathcal{R}^{1/2}}

\newcommand{\knoise}{k_{\mathrm{noise}}}
\newcommand{\kmax}{k_{\mathrm{max}}}

\newcommand{\mm}{\mathrm{mm}}
\newcommand{\mum}{\mu\mathrm{m}}
\newcommand{\ns}{\mathrm{ns}}
\newcommand{\mach}{\mathcal{M}}
\newcommand{\cs}{c_\mathrm{s}}

\newcommand{\snr}{\mathrm{SNR}}
\newcommand{\snrcrit}{\mathrm{SNR_{crit}}}

\DeclareUnicodeCharacter{2212}{-}

\begin{document}

\title{From 2D to 3D: Recovering Turbulent Density Dispersions from Noisy Data}

\author[0000-0001-8253-7407]{Luz L.~Jimenez Vela}
\affiliation{Florida State University, Physics, Tallahassee FL 32309, USA}
\email{luzljimvel10@gmail.com}

\author[0000-0002-0706-2306]{Christoph Federrath}
\affiliation{Research School of Astronomy and Astrophysics, Australian National University, ACT 2611, Australia}
\email[show]{christoph.federrath@anu.edu.au}

\author[0000-0001-6661-2243]{David C.~Collins}
\affiliation{Florida State University, Physics, Tallahassee FL 32309, USA}
\email{dccollins@fsu.edu}

\author[0000-0002-4808-7286]{Seth Davidovits}
\affiliation{Lawrence Livermore National Laboratory, Livermore, California, 94550, USA}
\email{davidovits1@llnl.gov}

\begin{abstract}
Turbulence plays a central role in shaping the structure and dynamics of the interstellar medium (ISM), governing the star formation rate (SFR) and the initial mass function (IMF). A key consequence of turbulence is the generation of density fluctuations, which regulate the amount of dense gas available for star formation. Accurate measurements of the three-dimensional (3D) turbulent density dispersion are therefore essential for understanding molecular-cloud structure and star formation. However, observations typically provide only two-dimensional (2D) column densities and are often affected by measurement/detector noise. The Brunt method estimates the 3D density dispersion from 2D column-density maps, but it does not account for finite signal-to-noise ratio (SNR). Here, we extend the method to recover the 3D turbulent density dispersion from noise-contaminated observations. Using numerical simulations spanning a range of density perturbation amplitudes and noise types, we identify a characteristic noise wavenumber, $\knoise$, corresponding to the intersection of the signal and noise spectra. Restricting the Brunt reconstruction to wavenumbers below $\knoise$ yields a denoised density-dispersion estimate that closely reproduces the noise-free result. We provide a practical prescription to determine $\knoise$ directly from the measurement SNR and image resolution. Alternatively, if the noise spectrum is known, it can be subtracted directly from the observed spectrum, eliminating the need to estimate $\knoise$. The proposed correction recovers the noise-free density dispersion with errors of $\lesssim5\%$ for $\snr\ge3$ and $\lesssim15\%$ for $\snr\ge1$, enabling substantially more reliable estimates of turbulent density fluctuations from noisy column-density data.
\end{abstract}


\section{Introduction} \label{sec:intro}

Turbulence plays a key role in star formation in the interstellar medium (ISM). On one hand, the turbulence creates overdensities in shocks on small scales, setting the seeds for gravitational collapse. On the other hand, it counteracts the effects of gravity on large scale, preventing the clouds from collapsing as a whole and thereby making star formation inefficient \citep[see, e.g.,][]{Larson81,ElmegreenScalo04,MacLow2004,McKee2007}. Turbulence may be driven by interstellar events such as supernova explosions, jets, galactic spiral arm dynamics, radiation from massive stars, and accretion, with the ability to create different mixtures of compressible gas motions, shearing motions, and vorticity \citep{Elmegreen09,Federrath17}.
 
Two main quantities that characterize turbulence are the density variance $\sigr$ and the sonic Mach number $\mach$. Their interrelation gives rise to the turbulence driving parameter, $b$, which is controlled by the turbulence driving mechanism, and takes the form
\begin{equation}
b = \sigr / \mach,
\end{equation}
where $\mach=\sigma_v/\cs$, with the standard deviation of the turbulent velocity fluctuations $\sigma_v$ and the sound speed $\cs$
\citep{Padoan97,Passot98,Price2011,Konstandin2012}. It is well documented that the driving parameter ranges from purely solenoidal (divergence-free) at $b=1/3$ to purely compressive (curl-free) at $b=1$, with a `natural mixture' represented by $b\sim0.4$ \citep{Federrath08,Federrath10}. The driving parameter plays a key role for the star formation rate (SFR) \citep{FederrathKlessen12} and the initial mass function (IMF) \citep{MathewFederrathSeta23}. Thus, we need reliable methods to measure $\sigr$ from observational data.

The key problem is that $\sigr$ quantifies the volumetric (3D) density fluctuations, while in observations we typically only have access to the projected (column; 2D) density fluctuations. Methods have been developed to estimate the 3D fluctuations from the 2D density variations \citep{Brunt10a,Brunt10b,Burkhart2012,KainulainenFederrathHenning2014,BruntFederrath14,YoonCho2024}, but these methods do not take into account the influence of the signal-to-noise (SNR) of the instrument/measurement. Thus, the goal of this work is to quantify the effects of the SNR for reconstructing $\sigr$ from observational data, and to provide a correction to account for finite SNR.

The paper proceeds as follows. Section~\ref{sec:methods} begins by presenting the simulation setup and analysis regions used to systematically quantify the effect of the SNR on $\sigr$ reconstruction. Section~\ref{sec:infinite} reviews the well-known case of $\mathrm{SNR}\to\infty$, accounting for different positions and different projections of the analysis region. The quantification of the dependence of $\sigr$ on SNR is presented in Section~\ref{sec:finite}. Section~\ref{sec:newbrunt} introduces the $\sigr$ reconstruction by a new denoising procedure. We conclude and summarize in Section~\ref{sec:conclude}.

\section{Simulations and basic methods} \label{sec:methods}

We use the simulations from \citet{Dhawalikar22}, which mimic a laboratory experiment of laser-induced, shock-driven turbulence conducted at the National Ignition Facility (NIF). The experiment takes place within a shock tube target for which these simulations have been made to aid in the understanding of hydrodynamic shock-driven turbulence \citep{Davidovits22,Dhawalikar22,Hew23}. Here we use it as a testbed for developing a new denoising technique. While the simulations and setup are designed for the experiment, the data taken are sufficiently general and representative of typical turbulent media, including the star-forming interstellar medium, that the new method developed here is applicable beyond the particular case studied.

\subsection{Hydrodynamic simulations} \label{sec:numericalsims}

The simulations involve a cylindrical tube (symmetry axis in the $y$ direction) with a total length of $3.8\,\mm$ and an outer diameter of $2.0\,\mm$ -- see details of the setup in \citet{Dhawalikar22}. The tube is filled with foam, and the foam is constructed to contain voids, which generate turbulence when the shock passes through the tube \citep[for possible experimental setups that can achieve this, see][]{Hamilton2016,Nagel17}. Here we compare 3~simulations that use foam voids with diameters of $12.5$, $50$, and $100\,\mum$. All materials are in pressure equilibrium at the start of the simulation. The laser beam is incident and initialized at the bottom of the shocktube along the $y$-direction inducing a plane-parallel hydrodynamic shock wave. Once the shock breaks out of the cylinder in all spatial directions, it is free to leave the computational domain, i.e., there is no influence of the numerical boundaries on the results.

The simulations use a modified version of the adaptive mesh refinement (AMR) code FLASH \citep{Fryxell00,Dubey2008,Federrath21} to solve the 3D compressible hydrodynamical equations. They use the robust HLL5R approximate Riemann scheme \citep{Waagan2011}. The computational grid has $768\times1024\times768$ cells, capturing the turbulent dynamics and the statistical properties of the turbulence as demonstrated in a sample of works \citep{Kitsionas2009,PriceFederrath2010_,Kritsuk2011}.

\begin{figure}
\centering
\includegraphics[width=\linewidth]{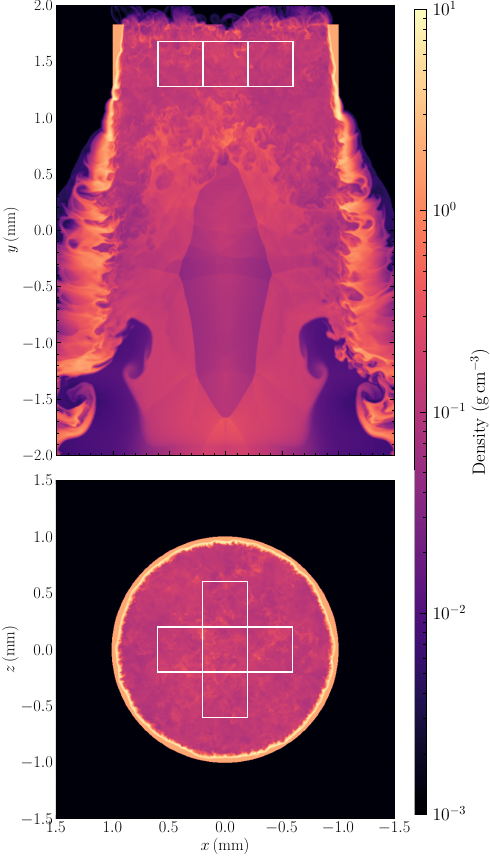}
\caption{Locations of the turbulence analysis regions (marked as white squares) in the shocktube simulation with $50\,\mum$ foam voids from \citet{Dhawalikar22}. Density slices at $z=0$ (top panel) and $y=1.5\,\mm$ (bottom panel) are shown at $t=75\,\ns$, i.e., just as the shock wave is exiting the tube at $y\sim1.8\,\mm$, leaving behind a turbulent post-shock region. Five different turbulence analysis regions are selected (see white rectangles), which cover regions of fully-developed, nearly isotropic turbulence. We compare these regions in order to quantify the dependence of our results on the particular choice of region.}
\label{fig:location}
\end{figure}

Figure~\ref{fig:location} shows slices through the material density in the simulation with $50\,\mum$ foam voids at the time when the shock wave is about to exit the tube (at $y\sim1.8\,\mm$), and there is a sufficiently large post-shock region of fully-developed turbulence. More details on the simulations can be found in \citet{Dhawalikar22}. For the present study, the details of the experimental setup are not crucial, since we extract information sufficiently far away from the boundaries.

\subsection{Turbulence analysis regions} \label{sec:analysisregion}

Five different turbulence analysis regions are selected (see Fig.~\ref{fig:location}), in order to study the variations of our results with choice of region. The dimensions of each region are $0.4\,\mm \times 0.4\,\mm \times 0.4\,\mm$, located in an analysis plane centered around the symmetry axis of the tube and at $y=1.5\,\mm$, where the post-shock turbulence is mostly isotropic and sufficiently far away from the edges of the cylinder. We use cubic regions to reconstruct the 3D density dispersion from 2D (projected) data \citep{Brunt10a}. We note that extensions of the \citet{Brunt10a} method to non-cubic domains have been explored recently in \citet{YoonCho2024}, and our results can be extended to such cases, accordingly.

\subsection{Calculating the intrinsic turbulent volume (3D) density dispersion}

Since our main goal is to develop a method to reconstruct the 3D density dispersion from projected (2D) column density data of finite SNR, we need to benchmark the reconstruction against the true (intrinsic) 3D turbulent density dispersion, which we denote as $\sigr$. To compute the latter, we normalize the density by its mean value ($\rho_0$) and calculate its standard deviation, $\sigr$, straight from the simulation data, as
\begin{equation}
\sigr = \left[\frac{1}{n} \sum_{i=1}^n (\rho_i/\rho_0 - 1)^2\right]^{1/2},
\label{eq:sigmarho3D}
\end{equation}
with the mean density 
\begin{equation}
\rho_0 = \frac{1}{n} \sum_{i=1}^n \rho_i,
\label{eq:rhoave}
\end{equation}
where $\rho_i$ is the density in each cell $i=1\dots n$ of a sub-region defined in Sec.~\ref{sec:analysisregion}, noting that each cell has the same size.
The values for the $50\,\mum$ void size case are listed in Table~\ref{tab:positions}.

\begin{table} 
\caption{Dependence on position and projection axis.}
\centering
\begin{tabular}{cccccc}
\hline
Position & Projection  & $\sigr$ & $\sign$ & $\mathcal{R}^{1/2}$ & $\sigrB$ \\
(1) & (2) & (3) & (4) & (5) & (6) \\
\hline
left & $x$ & 0.28 & 0.12 & 0.38 & 0.33 \\
right & $x$ & 0.26 & 0.11 & 0.39 & 0.28  \\ 
center & $x$ & 0.32 & 0.12 & 0.34 & 0.36 \\
back & $x$ & 0.30 & 0.12 & 0.34 & 0.36 \\
front & $x$ & 0.26 & 0.11 & 0.35 & 0.32 \\
\hline
center & $y$ & 0.32 & 0.11 & 0.31 & 0.35 \\
center & $z$ & 0.32 & 0.14 & 0.38 & 0.36 \\
\hline
\end{tabular}
\tablecomments{Comparison of (by columns): (1) different regions, and (2) projections, (3) their intrinsic density dispersion, (4) column density dispersion, (5) density dispersion ratio defined in Eq.~(\ref{eq:r}), and (6) the Brunt estimate of the density dispersion; Eq.~(\ref{eq:sigrB}), for the $50\,\mum$ foam void size for the five sub-regions shown in Fig.~\ref{fig:location}. The sample axis of projection is $x$ and the values for the $y$ and $z$ projections are listed for the center cube in the last two rows of the table.}
\label{tab:positions}
\end{table}

\subsection{Calculating the column (2D) density dispersion} \label{sec:bruntmethod}

To mimic observational data, such as those obtained via dust extinction or molecular line transitions in molecular clouds, or similarly those obtained via 2D radiographs in the laser experiments, we compute the 2D column density, $N$, of each analysis region by projecting the 3D density onto different 2D planes. For example, the projection along the $z$-axis is computed as
\begin{equation}
N(x,y) = \sum_j^{n_z} \rho(x,y,z_j) \Delta z
\end{equation}
where the sum goes over all cells $j=1\dots n_z$ along the $z$-axis, and $\Delta z$ is the cell length. Likewise, the two~other cartesian projections, $N(x,z)$ and $N(y,z)$ are obtained by summing over the $y$- and $x$-axis, respectively.

The column density dispersion, $\sign$, is obtained by replacing $\rho$ with $N$ and $\rho_0$ with $N_0$ in Eqs.~(\ref{eq:sigmarho3D}) and (\ref{eq:rhoave}). The values of $\sign$ for the three cartesian projections in the $50\,\mum$ void size and center cube case are listed in Table~\ref{tab:positions}.

\subsection{Reconstructing $\sigr$ via the Brunt method} \label{sec:alabrunt}

In order to reconstruct an estimate of $\sigr$ from $\sign$, \citet{Brunt10a,Brunt10b} developed a method involving the power spectrum of the column density, $\ptwod$, which is readily available via Fourier transformation of $N$, taking into account boundary effects \citep{Brunt2010,Brunt10a}. The core of the method assumes that the power spectrum of the 3D density field,
\begin{equation}
\pthreed(\rho,k) = \langle\vert\hat{\rho}\vert^2 4\pi k^2\rangle_{k=\vert{\bf k}\vert},
\end{equation} 
where $k$ is the wavenumber, $\hat{\rho}$ is the Fourier transform of $\rho/\rho_0$, and $\langle\dots\rangle_k$ denotes the average over spherical shells with radius $k=\vert{\bf k}\vert$ and thickness $dk$, can be approximated from the column density power spectrum,
\begin{equation}
\ptwod(N,k) = \langle\vert\hat{N}\vert^2 2\pi k\rangle_{k=\vert{\bf k}\vert},
\end{equation}
where $\hat{N}$ is the Fourier transform of $N/N_0$, and the average is over rings with radius $k=\vert{\bf k}\vert$ and thickness $dk$.

Moreover, from Parseval's theorem we know that
\begin{align}
\sigr^2 &= \int_{k=1}^{\kmax} \pthreed(\rho,k) dk \,\rightarrow\, \sum_{k=1}^{\kmax} \pthreed(\rho,k),\\
\sign^2 &= \int_{k=1}^{\kmax} \ptwod(N,k) dk \,\rightarrow\, \sum_{k=1}^{\kmax} \ptwod(N,k),
\end{align}
where the last step simply refers to summation in discrete $k$-space, given discrete column density data, i.e., an image with $N_\mathrm{pix}$ pixels on each side. For simplicity, we consider only images with unity aspect ratio here \citep[methods for treating arbitrary aspect ratios are discussed in][]{Brunt2010,Brunt10a}, such that each side of the image contains the same number of pixels, yielding a total of $N_\mathrm{pix}^2$ pixels. The maximum resolved wavenumber is then given by the Nyquist limit, $\kmax = N_\mathrm{pix}/2$ for even $N_\mathrm{pix}$ and $\kmax = (N_\mathrm{pix}-1)/2$ for odd $N_\mathrm{pix}$ (in units of the fundamental mode).

\citet{Brunt10a}'s main assumption is then that the 2D column density spectrum can be extended into 3D space to approximate the 3D density spectrum as 
\begin{equation}
\pthreed(\rho,k)\approx 2k\ptwod(N,k),
\label{eq:brunt_assumption}
\end{equation}
which yields the 'Brunt' ratio,
\begin{equation}
\mathcal{R} = \frac{\sign^2}{\sigr^2} = \frac{\sum_{k=1}^{\kmax} \ptwod(N,k)}{\sum_{k=1}^{\kmax} \pthreed(\rho,k)} \approx \frac{\sum_{k=1}^{\kmax} \ptwod(N,k)}{\sum_{k=1}^{\kmax} 2k\ptwod(N,k)},
\label{eq:r}
\end{equation}
where we have used Eq.~(\ref{eq:brunt_assumption}) in the last step. This defines the Brunt estimate of $\sigr$ as
\begin{equation}
\sigrB = \frac{\sign}{\rtohalf} = \left(\sum_{k=1}^{\kmax} 2k\ptwod(N,k)\right)^{1/2}.
\label{eq:sigrB}
\end{equation}
This is the core of the Brunt method, which states that given a column density image normalized by its mean, computing its power spectrum and summing over all relevant $k$ (noting that $k=0$ is excluded to yield the variance) weighted by $2k$, yields an estimate of the  3D density dispersion. \citet{Brunt10a} tested this method on a set of magnetohydrodynamical simulations of turbulence and showed that $\sigrB\approx\sigr$ to within a typical uncertainty of $\sim10-20\%$.

Table~\ref{tab:positions} provides $\rtohalf$ and $\sigrB$ for the $50\,\mum$ void case and the three cartesian projections of the center region. For comparison, in a fully isotropic turbulent media one often finds an average value of $\rtohalf=1/3$ \citep{Brunt10a}, and indeed, we find values comparable to that within variations by $10-20\%$. However, recent work shows that this may not always be the case \citep{YoonCho2024}, in particular if the dimension of the cloud in the line-of-sight direction is different from that in the perpendicular direction, with \citet{YoonCho2024} providing the respective correction for $\rtohalf$ in such cases. In the following we compare the Brunt estimate with the true value of $\sigr$ in the perfect-signal limit (infinite SNR).

\section{The case of infinite SNR} \label{sec:infinite}

Before quantifying the effects of the SNR for the reconstruction of $\sigr$ from $\sign$, we test whether position, projection axis, and/or foam void size have a particular influence on $\sigrB$.

\subsection{Dependence on position and projection axis} \label{sec:position}

Using the $50\,\mum$ foam void case as a representative example, we first test whether varying the specific position at the analysis region, as seen in the bottom panel of Figure~\ref{fig:location}, introduces significant variations in $\sigrB$. Computing $\sigrB$ for the 5~analysis regions, we find an average of $\sigrB=0.33\pm0.03$, where the standard deviation is used to quantify the variation across the 5~regions. Thus, there is relatively little variation ($\sim10\%$) between the regions (see Tab.~\ref{tab:positions}). Likewise, we test the variation for different projection directions, for the center region. We find $\sigrB=0.36\pm0.01$ across the 3~cartesian directions, showing variations of the order of $\sim5\%$ (see Tab.~\ref{tab:positions}). Therefore, we compute and report averages and variations of all analyses quantities across all 5~regions and all 3~projections. The typical variations for each quantity are about $10-20\%$.

\subsection{Dependence on foam void size}

\begin{table}
\caption{Dependence on foam void size (infinite SNR).}
\centering
\setlength{\tabcolsep}{4pt}
\begin{tabular}{ccccc}
\hline
Void size & $\sigr$ & $\sign$ & $\mathcal{R}^{1/2}$ & $\sigrB$ \\
(1) & (2) & (3) & (4) & (5) \\
\hline
$12.5\,\mum$ & $0.22^{+0.07}_{-0.03}$ & $0.10^{+0.02}_{-0.02}$ & $0.38^{+0.04}_{-0.04}$ & $0.27^{+0.03}_{-0.04}$ \\
$50\,\mum$ & $0.28^{+0.03}_{-0.02}$ & $0.12^{+0.01}_{-0.01}$ & $0.35^{+0.03}_{-0.02}$ & $0.33^{+0.03}_{-0.03}$ \\
$100\,\mum$ & $0.44^{+0.04}_{-0.05}$ & $0.20^{+0.05}_{-0.06}$ & $0.38^{+0.03}_{-0.04}$ & $0.52^{+0.09}_{-0.08}$ \\
\hline
\end{tabular}
\tablecomments{Same as Tab.~\ref{tab:positions}, but for different foam void sizes (col.~1), taking the median and its upper and lower limits denoting the 16th and 84th percentile variation (over all 5~positions (col.~2) and over all 5~positions and 3~projections (col.~3--5).}
\label{tab:voidsizes}
\end{table}

To quantify the dependence on the choice of initial perturbations as parameterized by the foam void size, in addition to the standard $50\,\mum$ voids, we study cases with $12.5\,\mum$ and $100\,\mum$ void sizes, respectively. Table~\ref{tab:voidsizes} lists the average values for all positions and all projection axes per foam void size for $\sigr$, $\sign$, $\rtohalf$, and $\sigrB$ respectively.

We see that the density fluctuations grow with increasing foam void size. This is expected, as stronger initial perturbations yield higher turbulent density fluctuations and Mach numbers after the shock passage \citep{Dhawalikar22}. For example, $\sigr$ and the Mach number increase by a factor of $\sim2$ between the $12.5\,\mum$ and $100\,\mum$ void cases. However, the Brunt $\rtohalf$ is relatively insensitive to these changes, displaying a certain level of universality across the different foam voids.

Finally, the reconstructed 3D density dispersion, $\sigrB$, provides an estimate of $\sigr$ that is slightly overestimated by $\sim20\%$, which is within the expected accuracy of the Brunt method \citep{Brunt10a}. Moreover, all reconstructed values, $\sigrB$, agree with the intrinsic $\sigr$ to within the 16th-to-84th percentile variations between the 15~different realizations (5~positions and 3~projection axes each), for each void size case, respectively.

\section{The case of finite SNR} \label{sec:finite}

Having confirmed the basic workings of the Brunt method in the limit of a perfect signal without noise, we are now ready to study the influence of a finite SNR of the instrument/measurement on the density dispersion reconstruction.

\subsection{Adding noise} \label{sec:noise}

\begin{figure*}
\centering
\includegraphics[width=\linewidth]{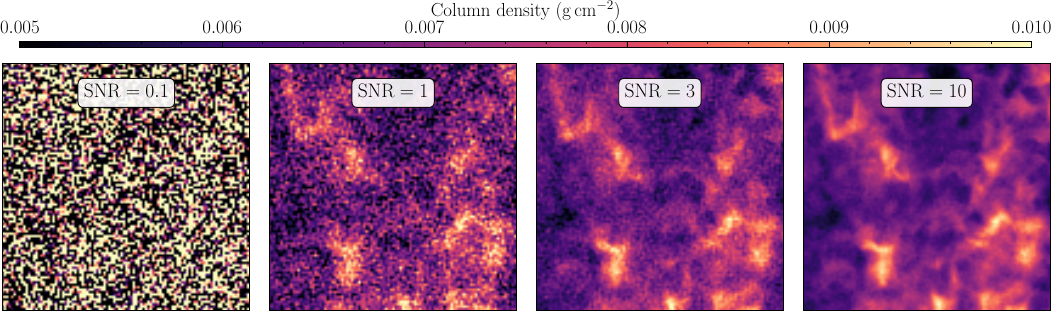}
\caption{Density projections along the $y$-axis of the center region with different levels of simulated white noise added -- from left to right: $\snr=0.1$, 1, 3, and 10. Noise dominates over the signal for $\snr\ll1$, while a nearly perfect signal is provided when $\snr\gg1$. Cases with $\snr\sim1$ are most interesting in that we can still reasonably recover the signal, but it may be strongly affected by noise, which we aim to correct for.}
\label{fig:foursnrs}
\end{figure*}

To mimic instrument/measurement noise, we begin by adding Gaussian (white) noise to the density-field projections of the simulations, to obtain 7~cases with the following SNR levels: 0.01, 0.1, 1, 3, 5, 10, and 100. The SNR is defined as the ratio of the intrinsic (pure signal) standard deviation of the column density field under consideration, divided by the standard deviation of the respective noise field, generated with the Fourier method outlined below. Figure~\ref{fig:foursnrs} shows examples of density projections with $\snr=0.1$, 1, 3, and 10. We consider these four to be of greatest interest both visually and for analysis purposes as we will show in the following sections.

Gaussian white noise has a flat power spectrum $\propto k^\beta$ with $\beta=0$. Noise with different $\beta$ can be generated by producing a spherically symmetric real-valued power law in Fourier space, $k^{\beta/2}$, such that when squared gives a power spectrum of $k^\beta$. The phase of each point in Fourier space is then randomized by multiplying each element by $e^{\mathrm{i}\theta}$, where $\theta$ is random and uniformly distributed in $[0, 2\pi]$. The inverse Fourier transform gives a random field with the target power-law noise spectrum,
\begin{equation}
\rm{Noise} = \mathcal{F}^{-1}\left(k^{\beta/2} e^{\mathrm{i}\theta}\right),
\end{equation}
where $\mathcal{F}^{-1}$ is the inverse Fourier transform operator. This noise signal is then a random field with the correct power spectrum.

For the 2D fields considered in this work, the corresponding azimuthally averaged spectrum scales as $P_{\rm{noise}} \propto k^{\beta+1} \propto k^\alpha$, due to integrating over $k$-shells with circumference $2\pi k$, and we choose to express and show the power-spectral noise exponents in terms of $\alpha$ instead of $\beta$. Appendix~\ref{app:typeofnoise} shows how the results depend on the distinct type of noise, where we compare the white-noise case ($\alpha=1$) with four other noise cases in which $\alpha=0$, $0.5$, $1.5$, and $2$, respectively.

\subsection{The effect of noise on the Brunt method}

\begin{figure}
\centering
\includegraphics[width=\linewidth]{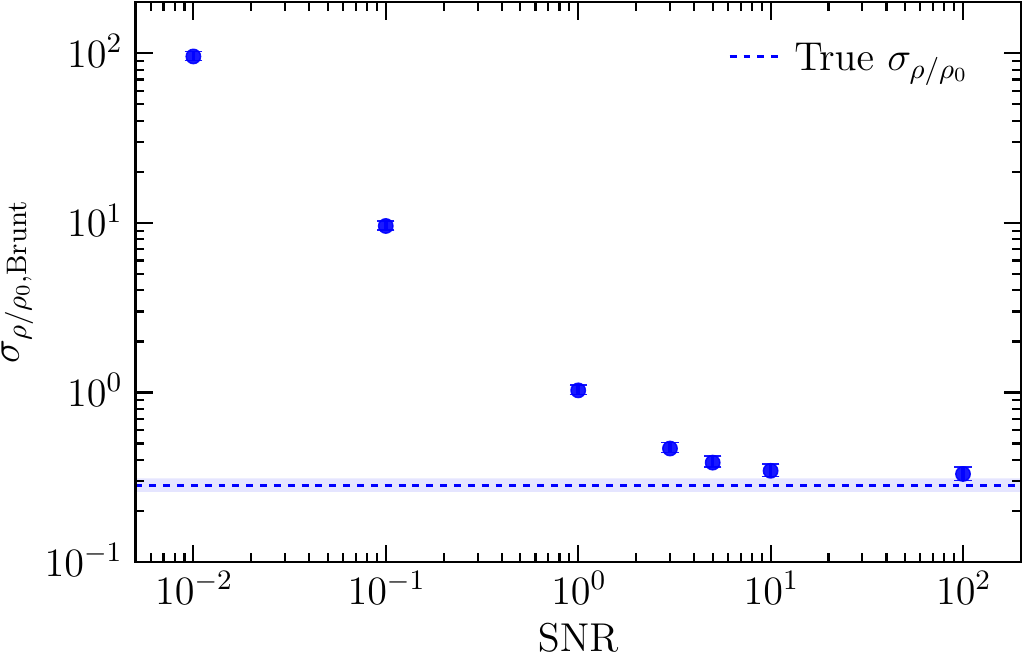}
\caption{Brunt density dispersion estimate $\sigrB$ as a function of SNR. The data points show the median and 16th to 84th percentile range over the 15~samples (5~positions with 3~projections each). As $\snr\to\infty$, the density dispersion reconstruction via the Brunt method works as expected, approaching the intrinsic $\sigr$ (shown as the dashed horizontal line with shaded area) to within $\sim20\%$ accuracy, while for $\snr\lesssim1$, $\sigr$ is strongly overestimated (by factors of several, up to orders of magnitude), requiring correction.}
\label{fig:noisysigmarho}
\end{figure}

Having added noise, we can now obtain the Brunt estimate, $\sigrB$, as described in Section~\ref{sec:alabrunt}, for each noise level, which is shown in Figure~\ref{fig:noisysigmarho}. First, we note that as $\snr\to\infty$, the $\sigrB$ estimate approaches the value from Sec.~\ref{sec:infinite}, as expected. With decreasing SNR, we see that $\sigrB$ strongly increases, ultimately overestimating the true $\sigr$ by $\sim2$~orders of magnitude at $\snr=0.01$. Thus, $\sigrB$ is completely dominated by noise in the limit $\snr\ll1$. However, even for $\snr\sim1$, the noise leads to an overestimate by factors of a few. Considering the overall uncertainties of $\sim20\%$ in the Brunt method in the case of noise-free data, biases by factors of a few constitute a major issue when considering measurements with $\snr\sim1$, which are commonly encountered in real observations and/or experiments. Thus, we need to develop a method to correct for this noise bias.

\section{A new noise correction method} \label{sec:newbrunt}

\subsection{Subtracting the noise spectrum} \label{sec:subtraction_method}
An ideal noise correction method would require complete knowledge of the noise, in which case the noise may be simply subtracted from the measured data, leaving pure signal. However, this is almost always impossible to do as the noise is an intrinsically stochastic process. However, one can sometimes apply a statistical approach in which at least an average noise spectrum may be measured, for example by taking measurements 'off target' or without any data target at all, which effectively provides a measurement of the typical noise encountered for a given instrumental setup. Indeed, if available, the average noise spectrum can then be subtracted from the measurement spectrum, which provides a very good noise-corrected $\sigrB$.

\begin{figure}
\centering
\includegraphics[width=\linewidth]{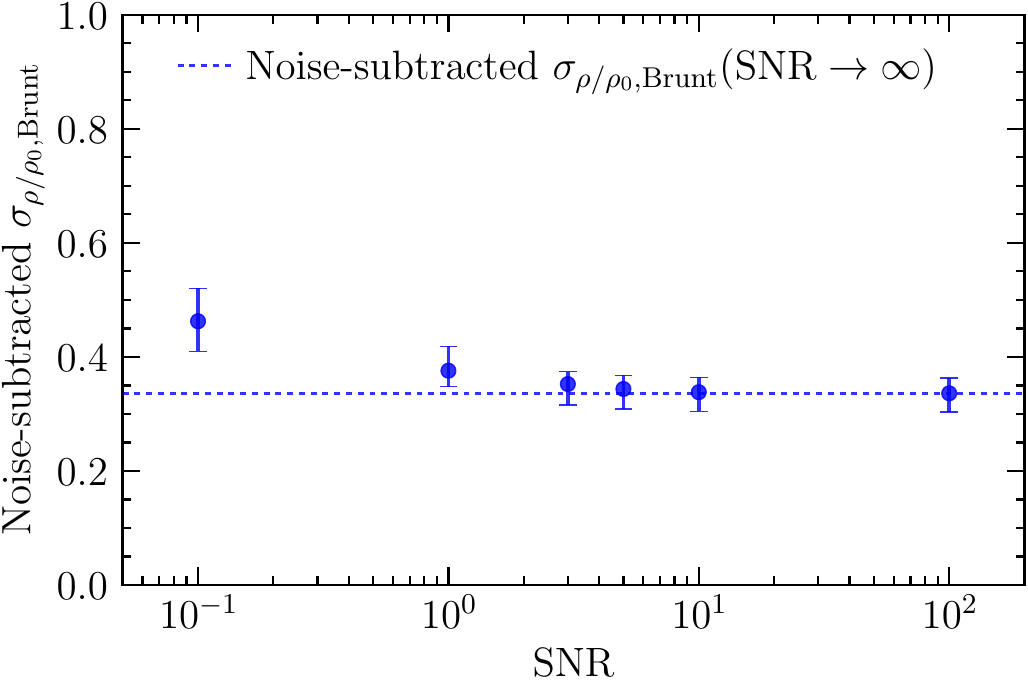}
\caption{Same as Fig.~\ref{fig:noisysigmarho}, but with the noise spectrum subtracted before evaluating Eq.~(\ref{eq:sigrB}).}
\label{fig:sigrB_noise_subtracted}
\end{figure}

We have tested this method on the same case as shown in Fig.~\ref{fig:noisysigmarho}. The result is shown in Fig.~\ref{fig:sigrB_noise_subtracted}, which plots $\sigrB$ obtained via subtraction of the noise spectrum as a function of SNR. We find that it matches the $\snr\to\infty$ limit for $\snr\ge3$ to within 5\% accuracy. Even at $\snr=1$, the overestimate is only $12\%$. This method would in principle work on any physical data (void size, Mach number, driving mode of turbulence, etc.). However, the downside of this method is that one requires the noise spectrum to perform the subtraction, but the noise spectrum may not be available. In that case, it is impossible to use this direct method. However, in the following, we develop an alternative, approximate method to correction for noise.

\subsection{Determining the noise-dominated scales} \label{sec:noise_scales}

\begin{figure}
\centering
\includegraphics[width=\linewidth]{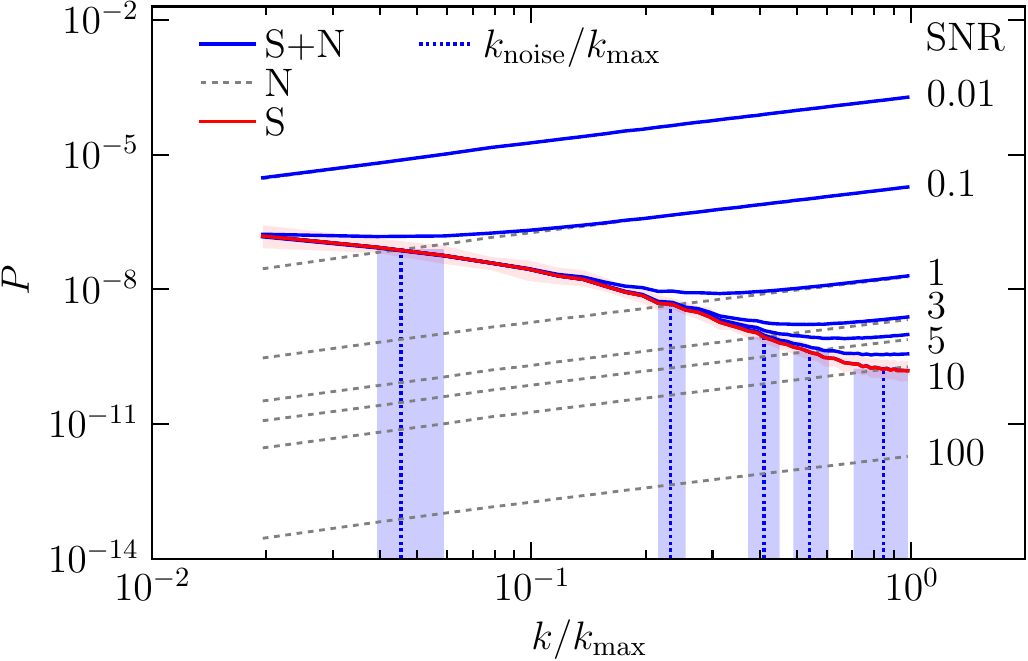}
\caption{Angle-integrated power spectra of the noise-contaminated column density (S+N, blue solid lines), the noise (N, gray dashed lines), and the original pure signal (S, red solid line), as a function of wavenumber $k$ in units of the maximum wavenumber $\kmax$ (set by the Nyquist frequency; c.f.~Sec.~\ref{sec:alabrunt}), for the 7~different SNR cases considered (as labeled on the respective curves/lines). The vertical blue dotted lines indicate the wavenumber $\knoise$, at which the signal and noise spectra intersect. The shaded regions demarcate the 15~samples' 16th-to-84th percentile ranges. The spectra are noise-dominated for $k\gtrsim\knoise$.}
\label{fig:spectra}
\end{figure}

To design a noise correction method for $\sigrB$ shown in Fig.~\ref{fig:noisysigmarho} that does not require detailed knowledge of the specifics of the noise (such as the noise spectrum), we first need to understand which scales are primarily dominated by noise. To do so, in Fig.~\ref{fig:spectra} we plot the angle-integrated power spectrum together with the power spectra of the noise, for different SNR. The noise-contaminated (signal plus noise: S+N) spectra are shown as blue solid lines, the pure noise (N) spectra are plotted as gray dashed lines, and the original, pure signal (S) is shown as the red solid line (with the shaded area delimiting the 16th to 84th percentile over the 15~realizations, as before). We see that the noise spectra follow the expected $\propto k^1$ scaling for white noise (c.f.~Sec.~\ref{sec:noise}), and their amplitude increases with decreasing SNR. The signal is a decreasing function of $k$, typical of turbulent density structures, including the ISM \citep[e.g.,][]{StutzkiEtAl1998_,SanchezEtAl2005_,KowalLazarianBeresnyak2007_,FederrathKlessenSchmidt2009_,FederrathKlessen2013_}.

We define the wavenumber where the noise and signal curves intersect as $\knoise$, shown as the vertical blue dotted lines. The intersection of N and S shifts to larger and larger scales (lower and lower $k$) as the SNR decreases. Thus, with decreasing SNR, more and more of the small-scale structures become noise-dominated. This $\knoise$ will form the basis for improving upon the existing Brunt method to estimate the density dispersion, given by Eq.~(\ref{eq:sigrB}). However, before we do so, we need to understand the properties of $\knoise$ in more detail.

\subsection{Properties of the noise wavenumber $\knoise$} \label{sec:knoise}

\begin{figure}
\centering
\includegraphics[width=\linewidth]{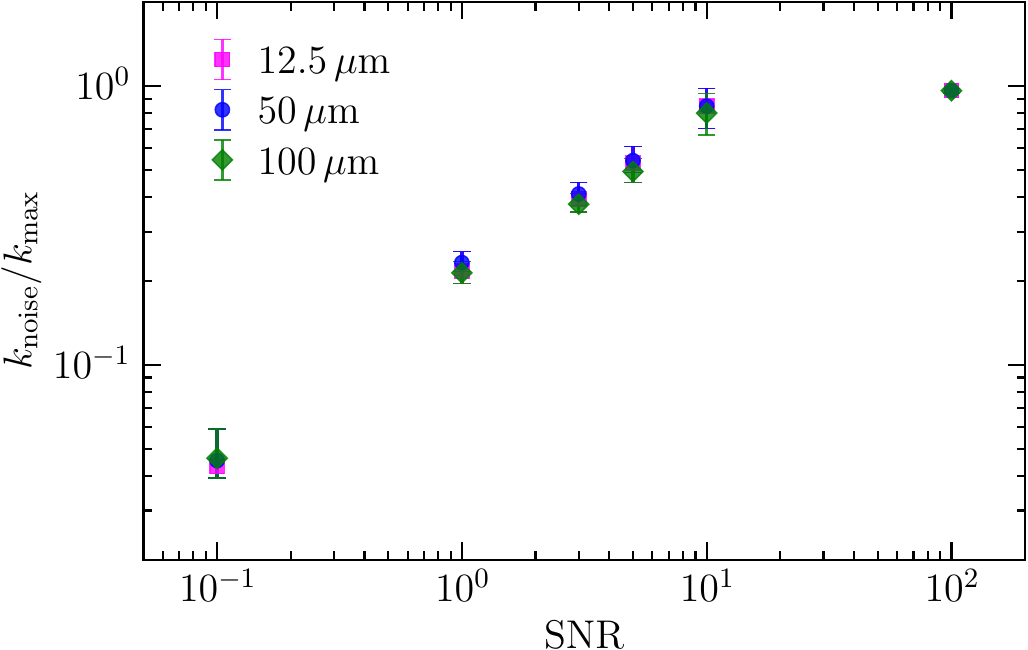}
\caption{The intersection wavenumber between the signal and noise spectra from Fig.~\ref{fig:spectra},  $\knoise$, as a function of SNR. In addition to the $50\,\mum$ void case (blue circles), the $12.5\,\mum$ (magenta squares) and $100\,\mum$ (green diamonds) void cases are shown as well. The error bars denote the 16th-to-84th percentile ranges of the 15~data samples. As expected, $\knoise\to0$ with decreasing SNR, while $\knoise\sim\kmax$ for $\snr\gtrsim10$, i.e., for high SNR, all scales can be considered, while for low SNR, smaller and smaller wavenumbers become noise-dominated. We see a universal dependence of $\knoise/\kmax$ on SNR, which forms the basis for the noise-correction method developed here.}
\label{fig:knoise}
\end{figure}

Figure~\ref{fig:knoise} shows $\knoise$ normalized to $\kmax$ as a function of SNR for the $12.5$, $50$, and $100\,\mum$ void size cases (the respective spectra for the 12.5 and $100\,\mum$ void size cases are shown in Fig.~\ref{fig:spectra_for_other_voids}). We see that $\knoise$ increases with SNR until it reaches $\kmax$ at $\mathrm{SNR}\to\infty$. In practice, this limit is nearly reached already for $\snr\gtrsim10$ for the present cases. We further observe a relatively universal behavior of $\knoise$ in that the dependence of $\knoise/\kmax$ on SNR is practically invariant to changes in the initial density seed (void size), and therefore in the sonic Mach number -- at least for the cases studied here, with $\mach=0.29$, $0.43$, and $0.64$ for the $12.5$, $50$, and $100\,\mum$ void size cases, respectively \citep{Dhawalikar22}. A caveat to this is that at much higher Mach number, we expect the density spectrum to change slope \citep{KimRyu2005,FederrathKlessen2013_}, which is expected to change the intersection of the signal and noise spectra. However, as quantified in \citet[][figs.~7 and 8]{FederrathKlessen2013_}, the change in slope of the density spectrum with Mach number, turbulence driving mode, or magnetic field strength in fully developed turbulent flows may be regarded as moderate in the context of the transition scale $\knoise$, although a detailed investigation is required to quantify the effect. We leave this for future work.

\begin{figure}
\centering
\includegraphics[width=\linewidth]{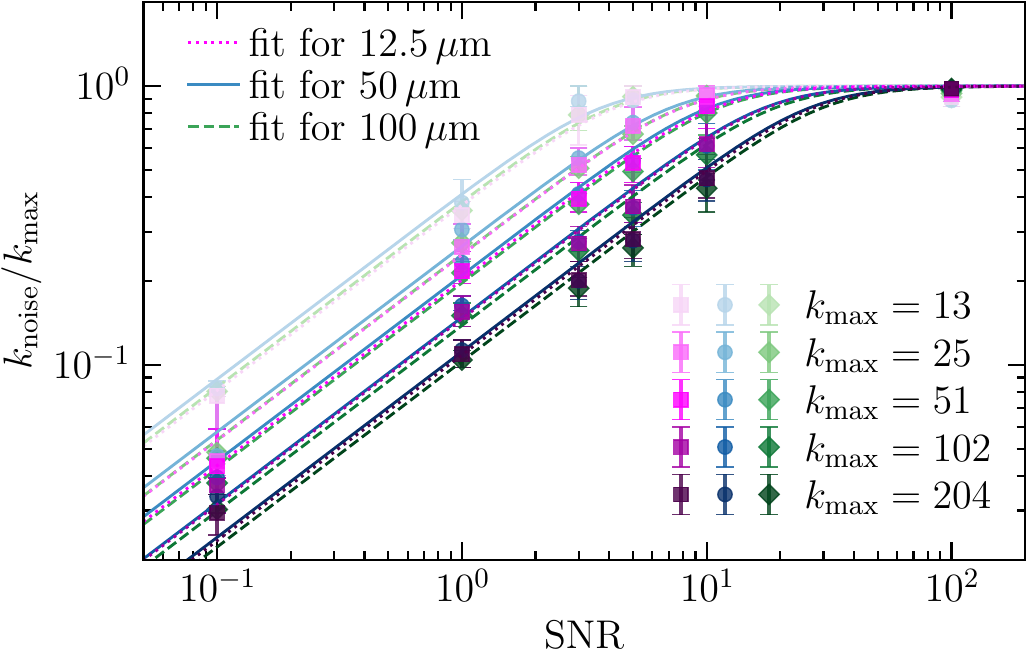}
\caption{Same as Fig.~\ref{fig:knoise}, but for resampled (interpolated) data with $2\times$ and $4\times$ reduced as well as increased resolution, respectively, leading to maximum wavenumbers of $\kmax=13$, 25, 51~(original), 102, and 204, for each of the three void-size cases, as indicated in the figure legend. The lines are fits using Eq.~(\ref{eq:knoise}), which determines $\snrcrit$ for the different $\kmax$ values.}
\label{fig:knoise_all}
\end{figure}

Another potential source of uncertainty in $\knoise$ is the resolution of the measurement/image, that is, the number of pixels, and therefore the maximum wavenumber $\kmax$ that is resolved in a measurement of the column density spectrum. The reason behind this is that the noise is concentrated on small scales. A detector that allows for higher spatial resolution, and therefore increased $\kmax$, is also subject to increased noise on the smallest scales. At fixed SNR, this means that the noise spectra shift down when $\kmax$ increases, resulting in a change in the intersection wavenumber $\knoise$. We quantify this effect by interpolating each original column density map to $2\times$ and $4\times$ higher and lower resolutions, respectively. The original maps extracted (c.f., Figs.~\ref{fig:location} and \ref{fig:foursnrs}) have $N_\mathrm{pix}=102\,\rightarrow\,\kmax=51$ (c.f., Sec.~\ref{sec:alabrunt}), leading to interpolated maps with $N_\mathrm{pix}=25,\,51,\,204,\,408\,\rightarrow\,\kmax=13,\,25,\,102,\,204$. The results for $\knoise/\kmax$ vs.~SNR are shown in Fig.~\ref{fig:knoise_all}, with example spectra for $\kmax=25$ and $102$ shown in Fig.~\ref{fig:spectra_res_resample}.

\begin{figure}
\centering
\includegraphics[width=\linewidth]{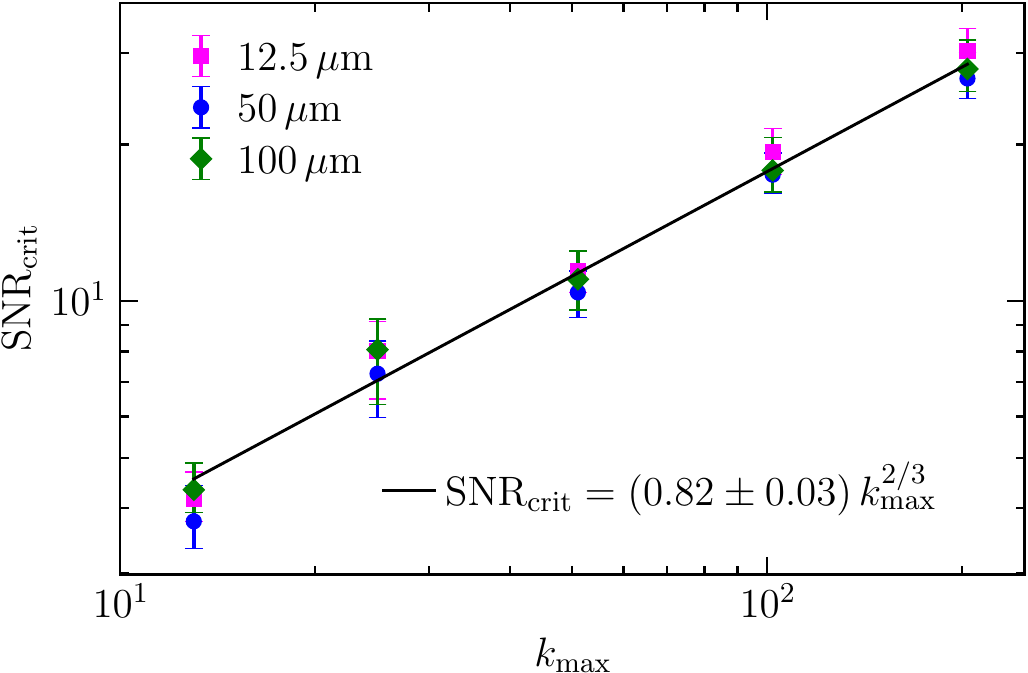}
\caption{The $\snrcrit$ in Eq.~(\ref{eq:knoise}) as a function of $\kmax$ for each of the three void-size cases. The line is a fit following Eq.~(\ref{eq:snrcrit}).}
\label{fig:snrcrit}
\end{figure}

Fig.~\ref{fig:knoise_all} shows that $\knoise$ depends not only on SNR, but also on $\kmax$. As before, the dependence on the void size is negligible. For each $\kmax$, we find that the data can be fitted well with the empirical function,
\begin{equation}
\knoise/\kmax = \left(\frac{1}{1 + (\snr/\snrcrit)^{-8/3}}\right)^{1/4}, \label{eq:knoise}
\end{equation}
where $\snrcrit$ is only a function of $\kmax$, representing a characteristic $\snr$, below which $\knoise\propto\snr^{2/3}$, and above which $\knoise/\kmax\to1$. Plotting $\snrcrit$ as a function of $\kmax$ in Fig.~\ref{fig:snrcrit}, we find that $\snrcrit$ itself follows a power-law relation with
\begin{equation}
\snrcrit = (0.82\pm0.03)\,\kmax^{2/3}. \label{eq:snrcrit}
\end{equation}
We have therefore determined the dependence of $\knoise$ on both $\snr$ and $\kmax$, with Eqs.~(\ref{eq:knoise}) and (\ref{eq:snrcrit}) providing an empirical model. While this model applies well in the case studied here, showing a certain level of universality with respect to the void size and therefore the strength of the turbulence (Mach number), we caution that it may not be fully universally applicable, as discussed above. Nevertheless, it may serve as a useful starting point for an informed noise correction to the original Brunt method, which we introduce now.

\subsection{The denoised density dispersion estimate} \label{sec:rtohalfdn} 

Now that we understand which scales are primarily affected by noise, namely wavenumbers $k\gtrsim\knoise$, we can recover the intrinsic power of the signal by filtering out those scales \citep[similar to the approach outlined in][]{AkahoriGaenslerRyu2014}, and only integrating the contaminated signal up to $\knoise$, which defines the denoised Brunt estimate in equivalence to Eq.~(\ref{eq:sigrB}) as
\begin{equation}
\sigrBdn = \left(\sum_{k=1}^{\knoise} 2k\ptwod(N,k)\right)^{1/2}.
\label{eq:sigrBdn}
\end{equation}

\begin{figure}
\centering
\includegraphics[width=\linewidth]{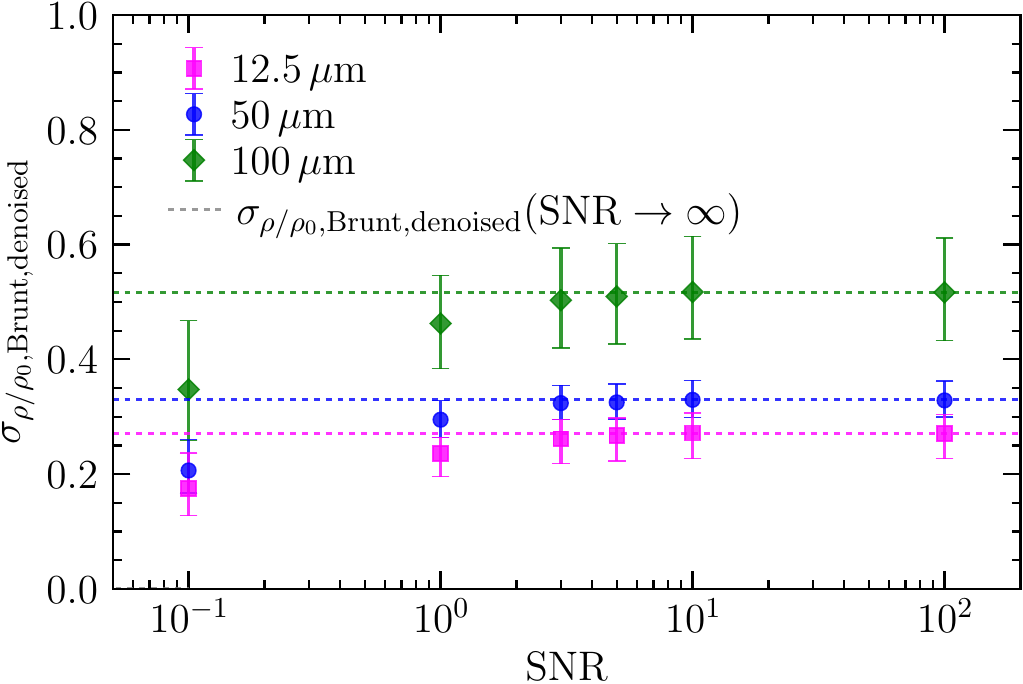}
\caption{Same as Fig.~\ref{fig:noisysigmarho}, but for the denoised density dispersion estimate, $\sigrBdn$, by evaluating Eq.~(\ref{eq:sigrBdn}), which uses $\knoise$ from Eqs.~(\ref{eq:knoise}) and~(\ref{eq:snrcrit}).}
\label{fig:sigrBdn}
\end{figure}

Fig.~\ref{fig:sigrBdn} shows the result of applying this noise correction, which can be directly compared to Fig.~\ref{fig:noisysigmarho}. In contrast to the noise-contaminated case, we see here that $\sigrBdn$ provides a very good estimate (accurate to within better than 15\%) of $\sigr$ for SNRs as low as $\sim1$. The error is only $\sim3-4\%$ for $\snr\ge3$, and still acceptably low with $10-13\%$ for $\snr=1$. Only for $\snr\lesssim1$ are the estimates starting to deviate significantly.

\section{Summary and conclusion} \label{sec:conclude}

We revisited the \citet{Brunt10b} method to estimate the volumetric (3D) density dispersion of a turbulent medium from projected (2D) column-density data. Such situations are frequently encountered in observations of interstellar clouds, where only line-of-sight integrated information is available, but also arise in terrestrial experiments where detector systems provide only integrated imaging data.

A key limitation of the original Brunt method is that finite signal-to-noise ratio (SNR) leads to systematically overestimated density-dispersion measurements (c.f.~Fig.~\ref{fig:noisysigmarho}). We therefore developed a new technique that accounts for finite-SNR effects and provides a denoised estimate of the 3D density dispersion.

Using simulation data representative of turbulent media found in the ISM and recent laser experiments, we analyzed five turbulence regions (c.f.~Fig.~\ref{fig:location}) and three different initial density perturbations to assess the robustness of the method. We first verified the applicability of the original Brunt method in the limit of a perfect signal ($\snr\to\infty$), and then added white and brown noise at SNR levels of 0.1, 1, 3, 5, 10, and 100.

By comparing noise-free, noise-contaminated, and pure-noise spectra (c.f.~Fig.~\ref{fig:spectra}), we identified a characteristic wavenumber, $\knoise$, corresponding to the intersection of the signal and noise spectra (c.f.~Figs.~\ref{fig:knoise} and~\ref{fig:knoise_all}). Restricting the Brunt reconstruction to scales $k\leq\knoise$ yields a denoised density-dispersion estimate via Eq.~(\ref{eq:sigrBdn}) that closely matches the $\snr\to\infty$ result (c.f.~Fig.~\ref{fig:sigrBdn}) for $\snr\gtrsim1$.

For practical applications, the denoised density dispersion can be obtained directly from Eq.~(\ref{eq:sigrBdn}) by summing only to $\knoise$. The required $\knoise$ follows from Eqs.~(\ref{eq:knoise}) and~(\ref{eq:snrcrit}) once the measurement SNR and $\kmax$ are specified. Alternatively, if the noise spectrum is known, it can be subtracted from the noise-contaminated spectrum before summation, in which case $\knoise$ is not required.

Although the calibration of $\knoise$ was obtained empirically from the simulations studied here, we found that it is largely insensitive to the amplitude of the density perturbations (Tab.~\ref{tab:voidsizes}) and to the type of noise considered (Appendix~\ref{app:typeofnoise}). Nevertheless, the density power spectrum is known to depend on additional physical parameters, including the turbulent Mach number \citep{KimRyu2005}, the driving mode of the turbulence, and, in self-gravitating systems, the evolutionary state of the cloud \citep[e.g.][]{KritsukNormanWagner2011,FederrathKlessen2013_,GirichidisEtAl2014}. Future work should therefore test the robustness of the present noise-correction method across a broader parameter space relevant to molecular-cloud observations.

Overall, the proposed noise-correction method recovers the $\snr\to\infty$ density dispersion with an error of $\lesssim5\%$ for $\snr\ge3$ and $\lesssim15\%$ for $\snr\ge1$, making the Brunt method applicable to substantially noisier datasets than previously possible.

\begin{acknowledgments}
\begin{nolinenumbers}
C.F.~acknowledges funding by the Australian Research Council (Discovery Projects grant~DP230102280 and DP250101526), and the Australia-Germany Joint Research Cooperation Scheme (UA-DAAD). C.F.~further acknowledges high-performance computing resources provided by the Leibniz Rechenzentrum and the Gauss Centre for Supercomputing (grants~pr32lo, pr48pi, and GCS Large-scale project~10391), the Australian National Computational Infrastructure (grant~ek9) and the Pawsey Supercomputing Centre (project~pawsey0810) in the framework of the National Computational Merit Allocation Scheme and the ANU Merit Allocation Scheme. S.D. performed work under the auspices of the U.S. Department of Energy by the Lawrence Livermore National Laboratory under Contract No. DE-AC52-07NA27344.
\end{nolinenumbers}
\end{acknowledgments}

\appendix

\section{Foam void sizes $12.5\,\mum$ and $100\,\mum$} \label{app:voidsizes}

\begin{figure}
\centering
\includegraphics[width=\linewidth]{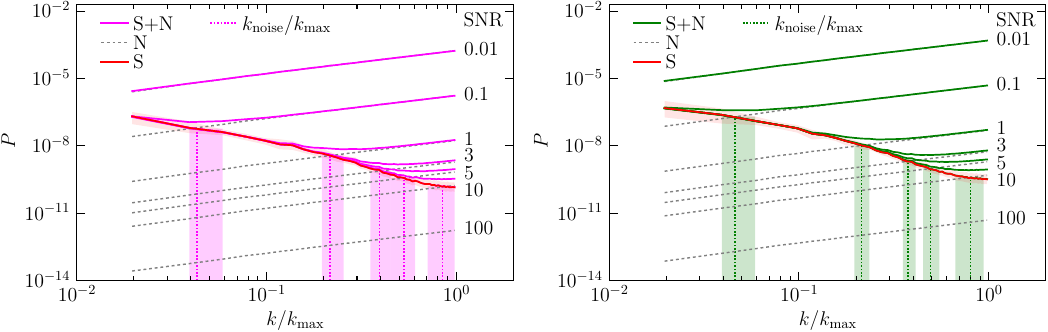}
\caption{Same as Figure~\ref{fig:spectra}, but for foam void sizes of $12.5\,\mum$ (left) and $100\,\mum$ (right).}
\label{fig:spectra_for_other_voids}
\end{figure}

Figure~\ref{fig:spectra_for_other_voids} shows the same as Fig.~\ref{fig:spectra}, but for foam void sizes of $12.5\,\mum$ (left) and $100\,\mum$ (right). We see that $\knoise/\kmax$ is nearly independent of the initial density perturbations (void sizes), which is explicitly shown in Fig.~\ref{fig:knoise}.

\section{Dependence on $\kmax$} \label{app:kmax_dependence}

\begin{figure}
\centering
\includegraphics[width=\linewidth]{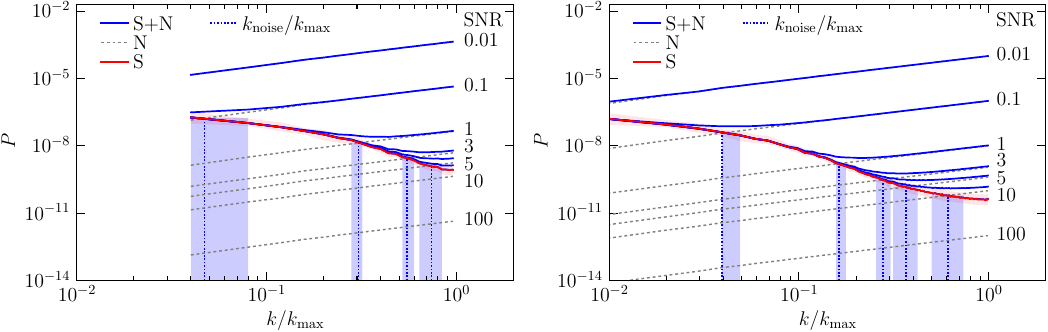}
\caption{Same as Figure~\ref{fig:spectra}, but for the resampled data with $2\times$ lower resolution ($\kmax=25$; left-hand panel) and $2\times$ higher resolution ($\kmax=102$; right-hand panel), respectively. As summarized in Fig.~\ref{fig:knoise_all}, we see that $\knoise$ depends on $\kmax$ as the noise is highest on the smallest scales (i.e., at $\kmax$).}
\label{fig:spectra_res_resample}
\end{figure}

Fig.~\ref{fig:spectra_res_resample} shows examples of the spectra similar to Fig.~\ref{fig:spectra}, but for the resolution-resampled maps to determine the effects of $\kmax$.

\section{On the type of noise} \label{app:typeofnoise}

\begin{figure}
\centering
\includegraphics[width=\linewidth]{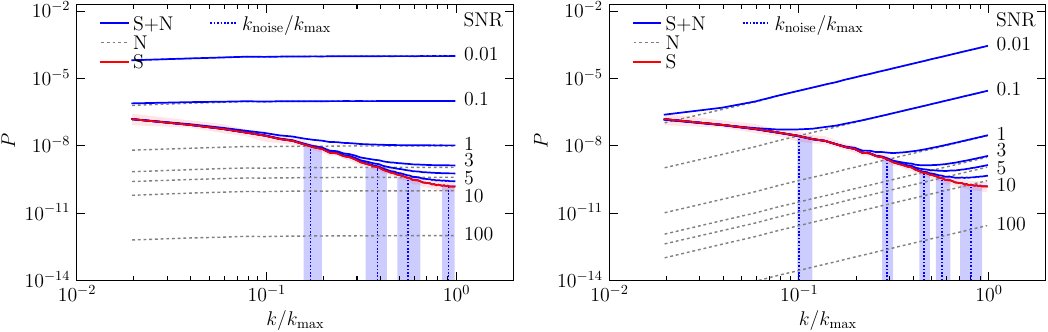}
\caption{Same as Figure~\ref{fig:spectra}, but for noise exponents $\alpha=0$ (left) and $\alpha=2$ (right) -- see Sec.~\ref{sec:noise} for the definition of $\alpha$.}
\label{fig:spectra_for_other_noise_exp}
\end{figure}

\begin{figure}
\centering
\includegraphics[width=0.5\linewidth]{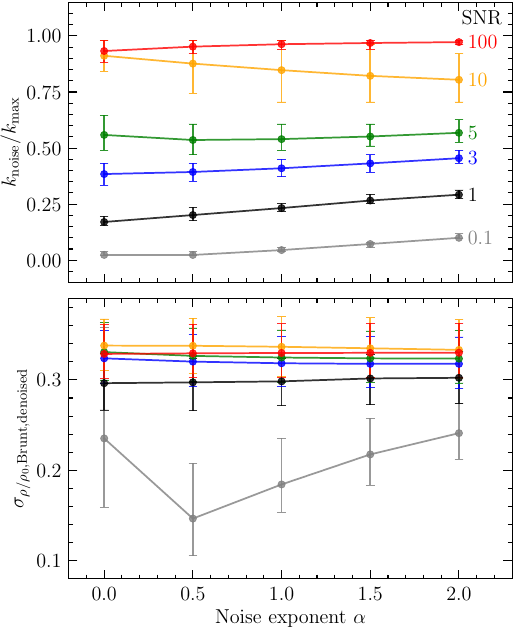}
\caption{Top panel: similar to Fig.~\ref{fig:knoise}, but showing $\knoise/\kmax$ as a function of the type of noise, parameterized by the noise exponent $\alpha$ (c.f.~Sec~\ref{sec:noise}), for different SNRs (as annotated to the right of each curve). Bottom panel: same as top panel, but for the noise-corrected Brunt estimate of the density dispersion, $\sigrBdn$, obtained via Eq.~(\ref{eq:sigrBdn}). We find that the noise correction method does not strongly depend on the noise exponent, as long as $\snr\gtrsim1$.}
\label{fig:knoise_for_other_noise_exp}
\end{figure}

Figure~\ref{fig:spectra_for_other_noise_exp} shows the same as Fig.~\ref{fig:spectra}, but for noise exponents of $\alpha=0$ (left) and $2$ (right), as generated via the methods described in Sec.~\ref{sec:noise}. We see that the noise spectra follow the target power-law exponent for $\alpha=0$ and $2$, respectively. Despite the significantly different shape of the noise spectra with varying $\alpha$, the intersection wavenumber ($\knoise$) with the signal spectrum, depends weakly on $\alpha$ for any given SNR, which is shown in the top panel of Fig.~\ref{fig:knoise_for_other_noise_exp}. The resulting de-noised Brunt estimate, $\sigrBdn$, is shown in the bottom panel of Fig.~\ref{fig:knoise_for_other_noise_exp}. We find that $\sigrBdn$ does not significantly depend on $\alpha$, except for $\snr\lesssim1$. We therefore conclude that the noise correction method is valid for different noise types as long as $\snr\gtrsim1$.

\end{document}